\documentclass[aps,pra,twocolumn,,superscriptaddress,floatfix]{revtex4-1}
\usepackage[svgnames]{xcolor}
\usepackage[colorlinks=true,urlcolor = purple,linkcolor=teal,citecolor=magenta,bookmarks=false]{hyperref}
\usepackage{amsfonts,amsmath,amssymb}
\usepackage{graphicx}
\usepackage{dsfont}

\usepackage{float}

\usepackage{tikz}
\usetikzlibrary{arrows,topaths,shapes.geometric,decorations.markings,shadows,positioning,
plotmarks}
 
\usepackage{pgfplots}

\begin{document}

\title{Domain wall melting in spin-1 chains}

\author{Marko Medenjak}
\affiliation
{Institut de Physique Th\'eorique Philippe Meyer, \'Ecole Normale Sup\'erieure, \\ PSL University, Sorbonne Universit\'es, CNRS, 75005 Paris, France}

\author{Jacopo De Nardis}
\affiliation{Department of Physics and Astronomy, University of Ghent, 
Krijgslaan 281, 9000 Gent, Belgium.}

\date{\today}

\begin{abstract}
In this letter we study the non-equilibrium spin dynamics in the non-integrable spin-1 XXZ chain, emerging after joining two macroscopic pure states with different magnetizations. Employing the so-called time-dependent variational principle (TDVP) we simulate the dynamics of the total magnetization in the
right (left) half of the system up to times that are normally inaccessible by standard tDMRG methods. We identify three distinct phases depending on the anisotropy of the chain, corresponding to diffusive, marginally super-diffusive and insulating. We observe a transient ballistic behaviour with a crossover time that diverges as the isotropic point is approached. We conclude that on intermediate-large time scales the dynamics is well described by the integrable Landau-Lifschitz classical differential equation. 
\end{abstract}


\maketitle

\paragraph*{\textbf{Introduction.}} 
One of the greatest quests of out-of-equilibrium statistical physics is obtaining an appropriate classical description of quantum non-equilibrium phenomena on hydrodynamical scales. While this has been achieved to large extent in integrable systems \cite{PhysRevLett.117.207201,PhysRevX.6.041065,PhysRevLett.120.045301,PhysRevB.96.115124,Bulchandani2017,Gopalakrishnan2018}, holographic theories \cite{Kovtun2003,PhysRevD.95.096003,Bhaseen:2015aa}, or at low temperature \cite{Sachdev19972,Damle1998,Damle2005}, it remains an open question when generic systems are under consideration. Nonetheless, it has recently been proposed that an appropriate description might be provided by the time-dependent variational principle (TDVP) \cite{Haegeman2016,PhysRevB.88.075133,Hallam2019,Leviatan2017} which, in contrast with standard tensor network approaches \cite{Paeckel2019} includes the conservation of local integrals of motion. The TDVP dynamics provides an effective non-linear classical time evolution on the space of matrix-product states (MPS) at fixed bond dimension and it is therefore expected to thermalize with similar qualitative \cite{Leviatan2017,TDVPProblems}, and in some cases also quantitative \cite{Leviatan2017}, features as the underlying quantum evolution of a generic quantum system. Such systems are expected to exhibit diffusive transport behaviour in general but it has recently been proposed that transport near-equilibrium might be altered even on intermediate-long time scales by the presence of global symmetries (despite the absence of integrability) \cite{DeNardis2019Anomalous,Richter2019,Dupont2018}.  This scenario was questioned in \cite{DupontMoore2019}, showing the necessity of probing the interplay between charge transport and global symmetries in a more controlled set-up.

 In this letter, we consider the transport properties of the non-integrable anisotropic spin-1 chain far from equilibrium. In particular we focus on two setups, the so-called domain wall and twisted initial conditions \cite{PhysRevE.59.4912,PhysRevE.71.036102,PhysRevE.78.031125,PhysRevE.78.061115,santos2009transport,lancaster2010quantum,PhysRevE.84.016206,PhysRevB.88.245114,Gamayun2019, Misguich2019,PhysRevB.99.121410}, which are experimentally relevant, and widely used for probing the transport and non-equilibrium dynamics in many-body systems \cite{Choi2016,PhysRevLett.115.175301,PhysRevA.92.053629,1906.06343}. We compute the time evolution using the TDVP algorithm at fixed bond dimension up to times  $\sim 10^3$, in units of spin coupling and compare it with the results obtained by a standard tDMRG algorithm. We observe that the time evolution of the time-integrated current, namely the total magnetization flowing from one side of the chain to the other, converges rapidly with the bond dimension employed in the simulation. 
 
The TDVP simulation with bond dimension $1$ corresponds to the classical mean-field evolution \cite{SciPostPhys.3.1.006}, while a larger bond dimension keeps the information about quantum correlations.  As we approach the isotropic point, we find that the quantum dynamics \textit{qualitatively} resembles the classical evolution in continuum space. Remarkably in the easy-plane regime $\Delta<1$ the spin transport is \textit{ballistic} up to a crossover time, which appears to diverge as $\Delta$ approaches $1^-$. In the isotropic case $\Delta=1$ the spin dynamics has the same qualitative behaviour as in the continuous classical \textit{integrable} Landau--Lifshitz (LL) partial differential equation \eqref{eq:LandauLifs} \cite{Faddeev1987,Takhtajan1977}, which indeed displays anomalous spin diffusion for non-twisted domain wall initial state \cite{Gamayun2019,Misguich2019}. In the easy-axis regime $\Delta>1$ we find \textit{insulating} spin dynamics for any value of the twist. 

Our results show that the large-time spin hydrodynamics of the domain wall initial state in a non-integrable chain close to its $SU(2)$ isotropic limit is \textit{qualitatively} described by the continuous Landau--Lifshitz  partial differential equation, which is integrable, up to extremely large times and it suggests that even in a non-integrable model one can observe anomalous spin transport up to the experimentally accessible times. While our work does not discuss standard linear-response spin diffusion constant at finite temperature, it points out that it might also be strongly affected by the corresponding classical dynamics.

\begin{figure*}[!t]
\center
\includegraphics[width=0.9\textwidth]{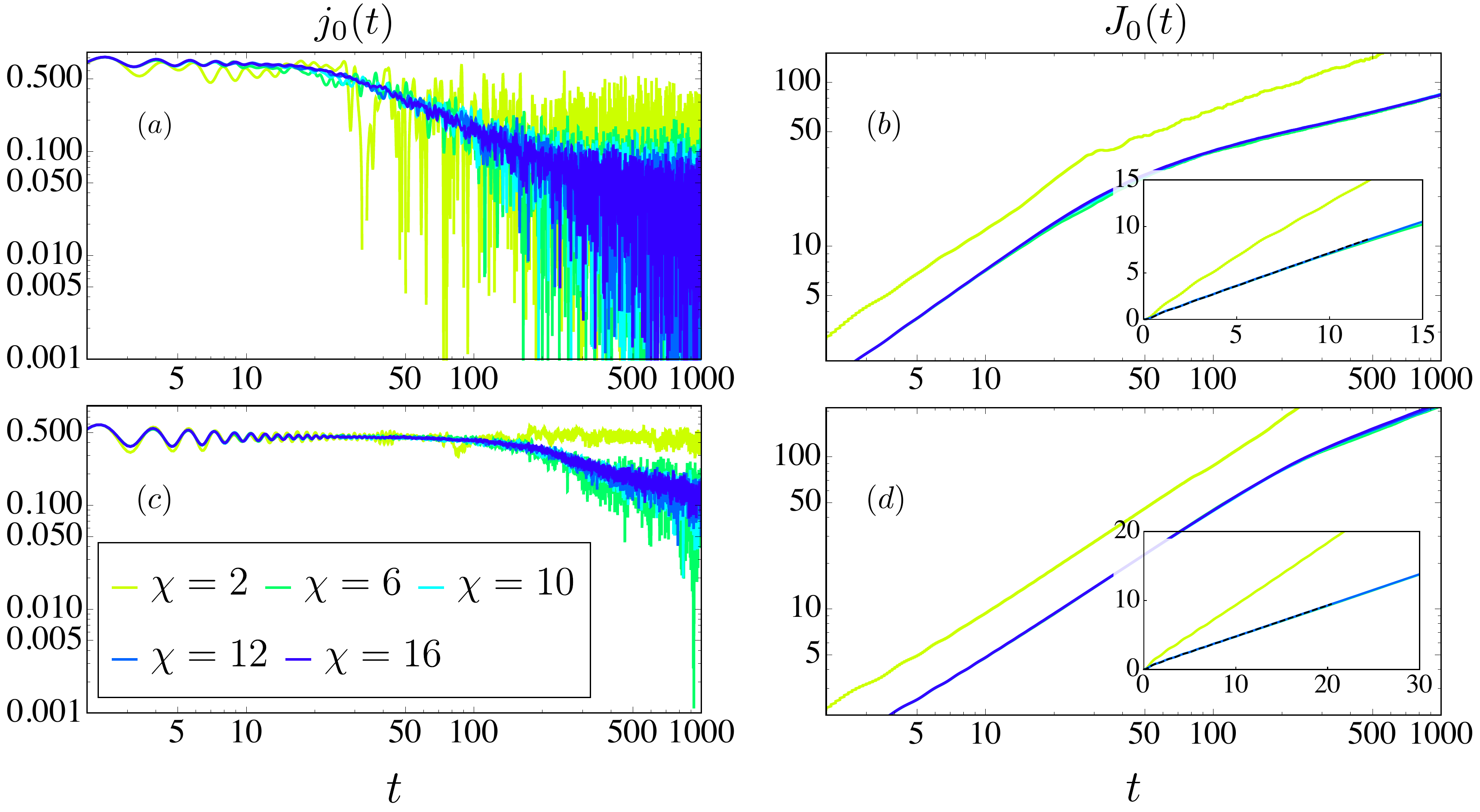}
\caption{ Left: Log-Log plot of the time evolution of the  local spin current $j_0(t) = \langle{  \rm DW}_\theta| \hat{j}_0(t) | {\rm DW}_\theta \rangle$ in the domain wall initial state with $\theta=\pi/4$ and $\Delta=0.5$ (a) and $\Delta=0.8$ (c) in the spin-1 chain, obtained by TDVP time evolution with bond dimensions $\chi$. We do not observe a clear convergence of the time evolution in bond dimension. Right: Log-log plot of the integrated spin current $J_0(t) = \int_0^t j_0(t') dt'$, for the same parameters as on the left, $\Delta=0.5$ (b) and $\Delta=0.8$ (d). A convergence of the time evolution of $J_0(t)$ in bond dimension for $\chi> 2$ can be observed at all times.  {Insets}: comparisons between data from TDVP time evolution and an exact TEBD evolution at short times.  }
\label{Fig:plot1}
\end{figure*}

\begin{figure}[t]
  \includegraphics[width=0.98\columnwidth]{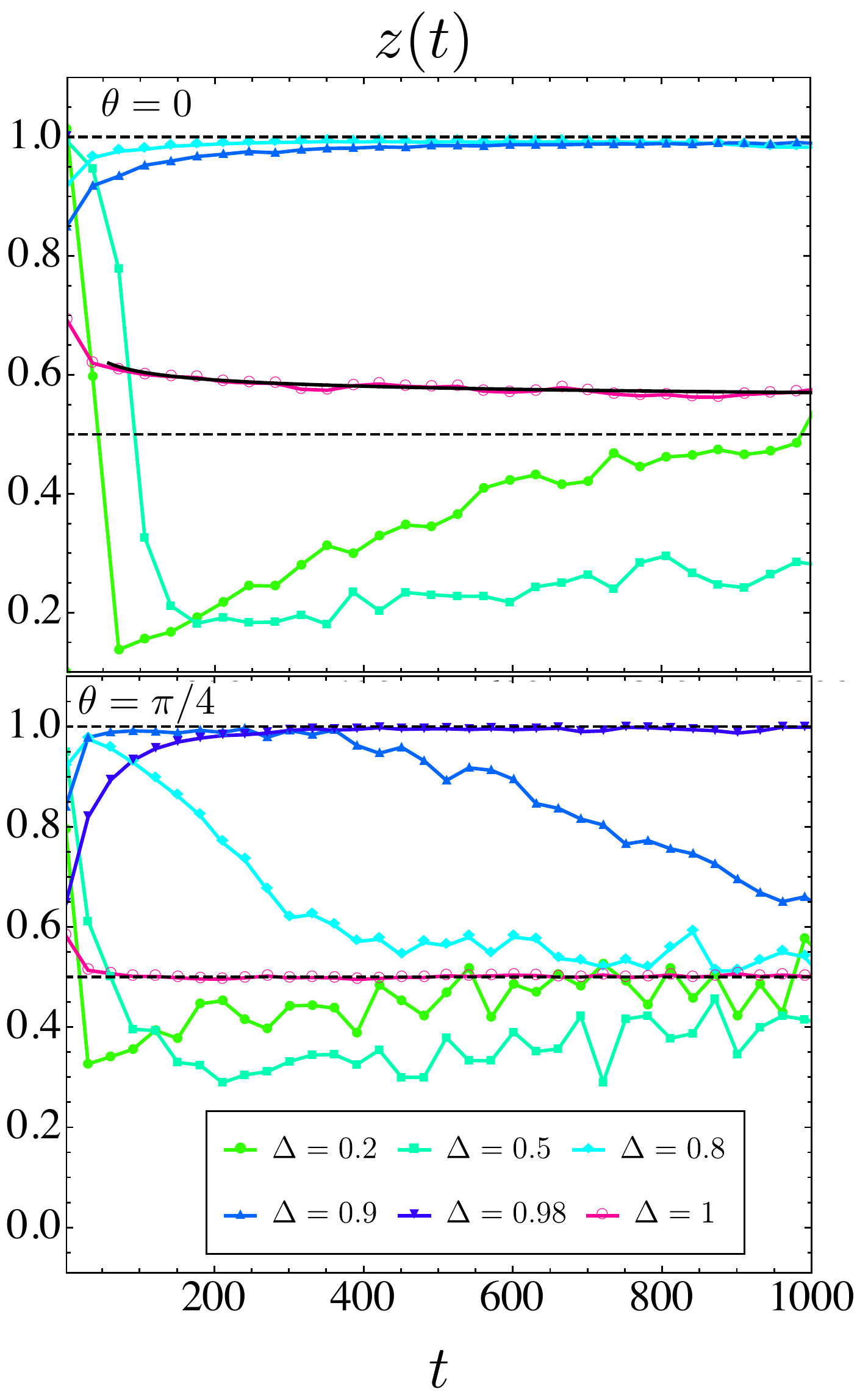}
  \caption{Top: plot of the exponent $z(t)$ obtained by dynamical fitting inside the interval $[t , t+34)$ for the domain wall initial state with $\theta=0$ for different values of $\Delta$ of the spin-1 chain with TDVP time evolution with $\chi=20$ (the function $z(t)$ is converged in bond dimension up to oscillations). We observe a crossover from ballistic to diffusive with a crossover time that increases as $\Delta \to 1^-$. At $\Delta=1$ we observe super-diffusive transport with log-corrections as in the LL classical dynamics up to arbitrary times (black line on top of $\Delta=1$ data is a fitting function $z(t)=1/2 + \kappa/\log t$, with $\kappa \simeq 0.48$. Bottom: the same data is displayed as in the plot on top but for the twisted domain wall initial state with $\theta=\pi/4$. We observe similar crossover from ballistic to diffusive transport for $ \Delta<1 $ and purely diffusive behavior at $\Delta=1$.   }
  \label{Fig:plot2}
\end{figure}

\paragraph*{ \textbf{The model and the domain wall quench.}}
In this letter we will consider the \textit{spin-1} XXZ Hamiltonian
\begin{equation}\label{eq:spinSchains}
\hat{H} = \sum_{j=-L/2}^{L/2} S^x_j   S^x_{j+1} +  S^y_j   S^y_{j+1}  + \Delta S^z_j   S^z_{j+1},
\end{equation}
where $S^{x,y,z}$ are the spin-1 representation of the $SU(2)$ spin operators, and $ \Delta $ is the anisotropy parameter. In order to study its non-equilibrium dynamics we consider a class of the tilted domain wall initial states
\begin{equation}\label{eq:initialState}
| {\rm DW}_\theta \rangle = \bigotimes_{j=-L/2}^{0} \left(  e^{- i \theta S_j^y} | \uparrow \rangle \right) \bigotimes_{j=1}^{L/2} \left(e^{ i \theta S_j^y} | \uparrow \rangle \right).
\end{equation}
We will mostly focus on the exact domain wall $\theta=0$ and on the case with $\theta=\pi/4$. In the first case the two halves of the chain are eigenstates, and the nontrivial dynamics is generated only around the junction at $j\sim 0$. This is no longer true, if $\theta\neq0$, and in this case the system is expected to thermalize to an equilibrium \textit{thermal} ensemble with the opposite local magnetization $ \langle S_j^z \rangle$. In order to study spin dynamics, we focus on the time-integrated local current $j_0 =  S_0^x S_1^y-  S_0^y S_0^x$
\begin{equation}
J_0(t) = \int_0^t dt' \langle {\rm DW}_\theta   |  \hat{j}_0 (t')| {\rm DW}_\theta   \rangle \sim t^{z(t)},
\end{equation}
which corresponds to the total magnetization transferred through the junction.
The domain wall quenches were extensively studied in the past years in the integrable spin-1/2 chain \cite{Calabrese2008,Mossel2010,Ljubotina2017,Misguich2017,Misguich2019,Gamayun2019}. Recently it was pointed out that the spin dynamics close to the isotropic point $\Delta - 1 = \delta  \sim 0$ is qualitatively described by the classical LL model \cite{Misguich2019,Gamayun2019},
\begin{equation}\label{eq:LandauLifs}
\partial_t \vec{S}_{x,t} = \vec{S}_{x,t} \times \partial_x^2 \vec{S}_{x,t} +   \vec{S} \times \bold{J} \vec{S},
\end{equation}
 where $\vec{S}_{x,t}$ is a classical three component vector field, normalized as $\vec{S}_{x,t}\cdot  \vec{S}_{x,t}=1$, and $\bold{J} = \text{diag}(0,0,\delta)$. In \cite{Gamayun2019} it was shown that this model exhibits a vast range of transport phenomena: transport is ballistic if $\delta<0$, and insulating if $\delta>0$. At the transient point $ \delta=0 $, the melting of the maximally polarized domain wall is marginally super-diffusive, and purely diffusive if $\theta\neq 0$. The late time behavior can be summarized with
 \begin{equation}
 J_0(t) \xrightarrow[t\to \infty]{ } \begin{cases}
\propto  t \text{  if  } \delta<0 \\
 \propto t^0  \text{  if  } \delta>0 \\
  \propto t^{1/2} (\log( b t))^d   \text{  if  } \delta=0  \text{  and  } \theta=0  \\
  \propto t^{1/2}  \text{  if  } \delta=0  \text{  and  } \theta\neq 0  
 \end{cases}
 \end{equation}
where $b,d>0$. 

In the quantum \textit{spin-1/2} chain a similar behaviour has been observed. Spin transport is indeed log-anomalous at $\Delta=1$ for a non-twisted domain wall \cite{Misguich2017,Stephan17}, and diffusive for a twisted initial conditions \cite{Ljubotina2017-diff}. In the $\Delta<1$ regime the transport is ballistic, which can be understood within the framework of the generalized hydrodynamics \cite{PhysRevLett.117.207201,CDV18}. While the ballistic transport is a consequence of quasi-local conserved operators (due to the integrability of the chain), which prevent the current from decaying \cite{Prosen11,ilievski2016quasilocal,PhysRevLett.111.057203,PhysRevB.96.020403}, the anomalous diffusion at $\Delta=1$ in the quantum chain is still lacking a theoretical explanation. 
 One of the major open questions related to transport in 1D systems, which we wish to address in this letter is therefore whether anomalous transport is merely a consequence of integrability, or if it is related to the global symmetry of the model and to its emergent classical hydrodynamical description.

\paragraph*{\textbf{Time evolution with MPS: TDVP time evolution.}}
As the model \eqref{eq:spinSchains} is non-integrable we have to resort to numerical simulations in order to compute its dynamics. The state of the art methods for probing dynamics of one-dimensional systems are based on MPS, which encode the information about the state locally, in terms of matrices $A^{s_i} \in \mathbb{C}^{\chi_{i-1} \times \chi_i}$ as
\begin{equation}
| \Psi[A] \rangle = \sum_{\{ s_{-L/2}, \ldots , s_{L/2} \}}^d A^{ s_{-L/2}} \ldots A^{ s_{L/2}} |  s_{-L/2}, \ldots , s_{L/2} \rangle,
\end{equation}
where $ d =3$ corresponds to the dimension of the local Hilbert space.

Initially the system is in a product state, $\chi = \chi_i =1$, however, due to the rapid growth of entanglement entropy $S_{\rm EE}(t)$, the bond dimension $ \chi_i \sim \exp S_{\rm EE} $ should be increased exponentially with time in order to keep the error under control \cite{PhysRevLett.93.040502}. Accordingly to the ideas of hydrodynamics, most of the information about the state is irrelevant for the large scale description of the charges and their currents, so one might attempt to capture the correct hydrodynamics by fixing the bond dimension $\chi$, and keeping only judicially chosen parts of information about the quantum correlations \cite{Leviatan2017}. This can be accomplished efficiently by employing the TDVP algorithm \cite{Haegeman2011,Haegeman2016,PhysRevB.88.075133}, whose time evolution is given by
 \begin{equation}
\frac{d | \Psi[A ] \rangle}{dt} = - i P_{\mathcal{M}_\chi} \hat{H}  | \Psi[A ] \rangle,
\end{equation}
where $P_{\mathcal{M}_\chi}$ is the projector to the manifold of MPS with a fixed bond dimension $\chi$. In contrast with other MPS--based methods TDVP preserves energy and single--site integrals of motion. The approach was used to compute linear response dynamics in quantum chains  at infinite temperature \cite{Leviatan2017}, and it required an ensemble average over initial conditions. In \cite{TDVPProblems} it was shown that the method captures the dynamical exponents of non-integrable (and non-free) spin chains precisely, however, it might fail to reproduce the correct \textit{quantitative} values of the transport coefficients, i.e. diffusion constants. Here we focus on the dynamics of a single initial state, and argue that the TDVP captures the qualitative features of the time-integrated spin current in the non-integrable chain \eqref{eq:spinSchains}. While the value of the current $ j_0(t) $ depends strongly on the bond-dimension $\chi$, this is not the case for the time--integrated  $j_0(t)$, see Fig. \eqref{Fig:plot1}. {This is a consequence of the ergodicity of the chaotic trajectories \cite{Hallam2019} in the phase space of MPS wave functions. While any single trajectory depends strongly on the bond dimension, the relaxation to the phase space average that we are probing with the time integration does not. }


\begin{figure}[t]
  \includegraphics[width=\columnwidth]{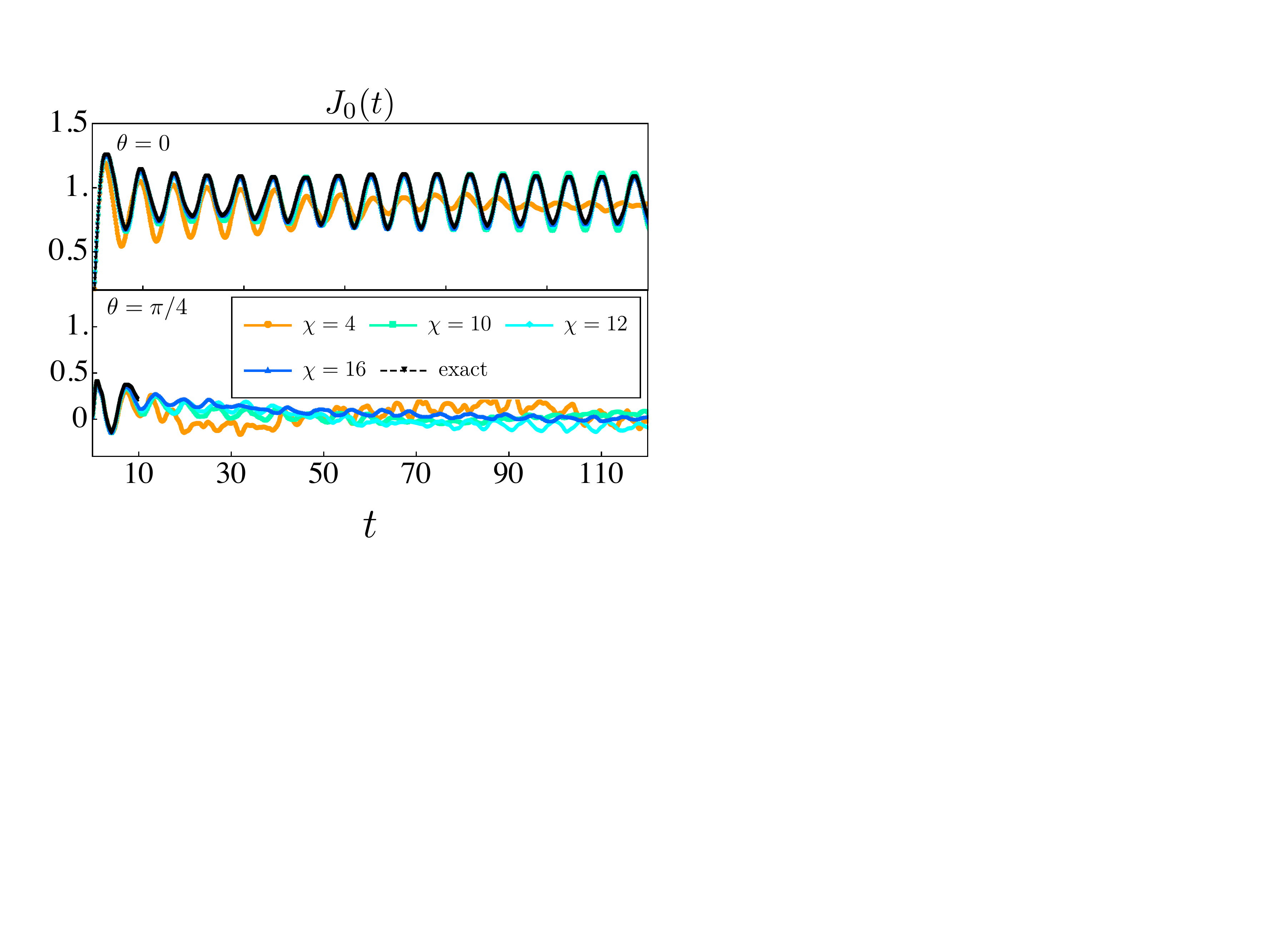}
  \caption{ Plot of the time-integrated current $J_0(t)$ for the spin-1 chain with $\Delta=1.5$ using TDVP time evolution with different bond dimensions $\chi$ and with the exact TEBD evolution. 
  Top:  domain wall with $\theta=0$. We observe insulating behaviour with periodic oscillations that converge with small bond dimension.   Bottom: for the domain wall with $\theta=\pi/4$ we again observe insulating behaviour for any bond dimension, although the profile does not show a clear convergence in bond dimension $\chi$.}
  \label{Fig:plot3}
\end{figure}

\paragraph*{\textbf{Main results.}}


The TDVP dynamics for the bond dimension $\chi=1$ corresponds to the classical mean field evolution. If we consider  the Hamiltonian \eqref{eq:spinSchains}, the time-evolution coincides with that of a classical Heisenberg chain
$
H_{\rm cl} = \sum_j \vec{S}_j \cdot \vec{S}_{j+1},
$
where $\vec{S}_j = (\sin \theta_j \cos \phi_j, \sin \theta_j \sin \phi_j, \cos \theta_j)$ is the 3-dimensional spin vector at site $j$, which can be equivalently represented by the product state
$
| \Psi_{\chi=1} \rangle =  \bigotimes_{j=-L/2}^{L/2} \left(  e^{- i \phi_j S_j^z} e^{- i \theta_j S_j^y} | \uparrow\rangle \right).
$
 Let us stress that the classical Heisenberg model is non-integrable, however its continuous limit corresponds to the integrable LL equation \eqref{eq:LandauLifs}. Increasing the bond dimension results in more complicated classical Hamiltonian dynamics, and the numerical results suggest that the qualitative transport behaviour is converged already using a very small bond dimension  $\chi \sim 6$. 
 
In Fig. \eqref{Fig:plot2} we plot the dynamical exponent $z$ for the growth of the time-integrated current for $\theta=0$. In the regime $\Delta< 1$ we observe a \textit{sharp} crossover from ballistic $z=1$, to a transient sub-diffusive regime with $z<1/2$, which is compatible with the restoration of normal diffusion $z=1/2$ at later times. As $\Delta$ approaches $1$ from the left, the crossover time increases, and we observe no deviation from ballistic behaviour up to times of order $10^3$ for $\Delta=0.8$ and $\Delta=0.9$. At $\Delta=1$ we find anomalous diffusion with logarithmic corrections, namely $J_0(t) \sim \sqrt{t} (\log (b t))^d$ with $d,b>0$ up to the maximal times considered. In Figure \ref{Fig:plot2} the dynamical exponent $z(t)$ are obtained with numerical simulations with bond dimension $\chi=20$, but the qualitative features of the profile are independent of $\chi$ for values larger than $\chi \sim 4$. Interestingly the TDVP time evolution at any maximal bond dimension shows the same transport dynamics as the classical integrable LL equation up to crossover times that diverge as we approach the isotropic point. A heuristic explanation of this behaviour at $\theta=0$, based on the energy conservation was offered in \cite{Misguich2019}. Initially, energy is concentrated around the link $j=0$. Due to the finite value of energy in the system, neighbouring spins are mostly aligned, making the quantum dynamics effectively classical and regulated, up to very large scales, by the LL equation  \eqref{eq:LandauLifs}. 
 
\begin{figure}[t]
  \includegraphics[width=\columnwidth]{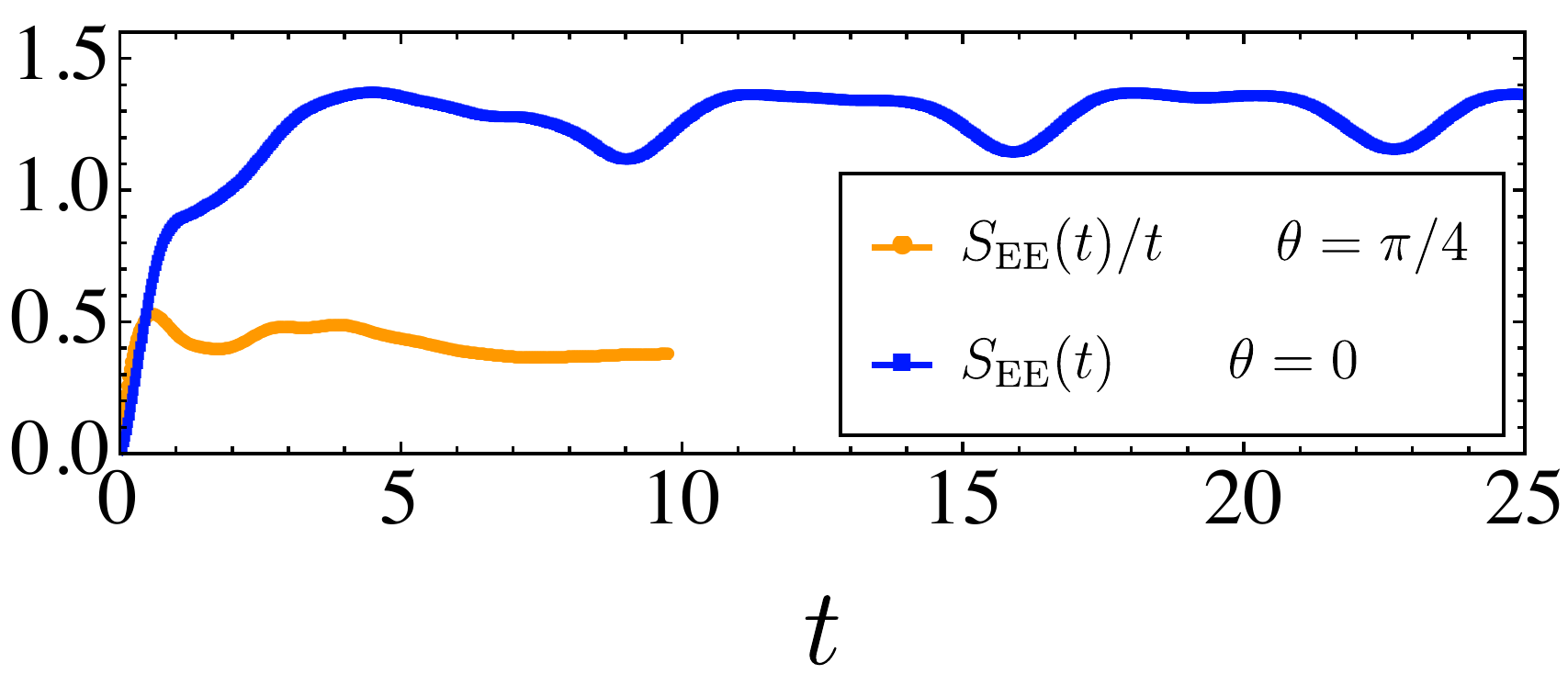}
  \caption{ Plot of the entanglement entropy evolution $S_{\rm EE}(t)$ after the domain wall quench with $\Delta=1.5$ obtained with exact TEBD simulation for $\theta=0$ and $\theta=\pi/4$. In the first case we observe saturation of entropy with periodic oscillations, while in the second case entropy increases linearly with time (we observe the same behavior for $\Delta<1$ and $\theta=\pi/4$). }
  \label{Fig:plot4}
\end{figure}

If $\theta=\pi/4$ (and in general $\theta\neq0$) the two halves are no longer eigenstates of the chain for $\Delta\neq 1$, and each side of the chain undergoes non-equilibrium time evolution. One would expect that two sides of the chain far from the junction thermalize, and that the system displays normal diffusion. We observe, however, similar behaviour as for the pure domain wall $\theta=0$, where the spin transport becomes ballistic as $\Delta$ approaches the isotropic value. For any $\Delta < 1$ the initial state has finite energy density throughout the chain, which makes the argument from \cite{Misguich17} inapplicable. While the crossover time is shorter than in the case of $\theta=0$, one can still observe ballistic dynamics up to very large times, $O(10^2)$, even for $\Delta\sim 0.8$.
 Instead at $\Delta=1$ we find normal diffusion in analogy with the classical LL evolution in the presence of a finite twist $\theta$. 
Interestingly, for $\Delta>1$ the transport is insulating both for the pure and the twisted domain wall, see Fig. \ref{Fig:plot3}. However, while we observe the convergence of the integrated current with bond dimension for $ \theta=0 $, the results do not seem to converge for the twisted initial conditions. 
Interestingly the domain wall $\theta=0$ shows persistent periodic oscillations of magnetization, see Fig. \ref{Fig:plot3}, and also of entanglement entropy, see Fig. \ref{Fig:plot4}. This behavior is reminiscent of scarred quantum states \cite{PhysRevLett.122.220603}, where in particular, periodic orbits in the semi-classical TDVP time evolution can be identified \cite{PhysRevLett.122.040603}.  While in the classical model the absence of transport is related to the presence of static soliton solutions \cite{Gamayun2019,Misguich2019} and their ``breathing'' modes, our analysis suggests that even the non-integrable spin-1 quantum chains posses eigenstates which have a large overlap with semi-classical localized states. On the other hand the case with $\theta=\pi/4$ shows linear growth of entanglement, see Fig. \ref{Fig:plot4}, despite the absence of transport. We are however unable to determine whether TDVP results are correct at large times in this case, since there is no convergence of the integrated current with the increasing bond dimension. This is likely related to the lack of ergodicity in the MPS phase space and not to the entanglement growth, since the same linear growth is observed also for $\Delta<1$, where the integrated current converges with the bond dimension.

 \paragraph*{\textbf{Conclusions and outlook.}} 
We have studied the dynamics of tilted and non-tilted domain wall state under the unitary evolution generated by the XXZ spin-1 chain. We have employed TDVP time evolution which shows rapid convergence of the dynamics of total magnetization on one half of the chain with increasing bond dimension.  Close to the isotropic point the TDVP dynamics gives qualitatively the same behaviour as the LL integrable partial differential equation of motion. Despite the non-integrability of the chain the domain wall melts with logarithmic super-diffusion for all times in isotropic chain and exhibits ballistic behaviour in the easy-plane regime up to very large time, which seems to diverge as the isotropic point is approached. In the easy axis regime, the system exhibits insulating behaviour even in the case of twisted domain wall, where one would naively expect diffusive transport.  Our work confirms that a \textit{non-integrable} quantum lattice model can be described by an \textit{integrable}  classical model and that the integrability breaking effects can be absent for arbitrarily large times. 

An interesting future perspective is to relate the time scales regulating the transition from ballistic to diffusive transport in the $\Delta<1$ regime, to the other thermodynamic quantities at play (energy density) and the anisotropic parameter $\Delta$. Moreover the sharp transition between ballistic dynamics to the intermediate sub-diffusive regime that we observe here for the domain wall state is also an interesting unexplained phenomena. A relevant question is if such a transition always occurs if the $SU(2)$ symmetry is perturbed.  Finally the mechanism regulating the insulating behaviour, and saturation of entanglement entropy in the quantum chain for $\Delta>1$ remain to be uncovered.

Here we restricted our discussion to the out-of-equilibrium spin dynamics arising from a single initial state, however the nature of the transport at finite temperatures remains controversial \cite{DeNardis2019Anomalous,Richter2019,Dupont2018,DupontMoore2019}. Our results suggest that close to the isotropic point the dynamics of the quantum non-integrable chain may be well described by its classical counterparts, which is expected to show anomalous super-diffusive dynamics \cite{Bojan,PhysRevE.100.042116} up to very large times.  It seems that the TDVP time evolution is a promising candidate to answer these questions. 

 \paragraph*{\textbf{Acknowledgements.}} 
The MPS-based TDVP and TEBD simulations were performed using the ITensor Library  \cite{ITensor}. Special thanks to Enej Ilievski for collaborations in related works and to Maarten van Damme for early collaboration on the work and to Jutho Haegeman and Andrew Hallam for helping with the development of a TDVP code and for numerous insightful discussions. J.D.N. is supported by Research Foundation Flanders (FWO).

\bibliography{LowT}

\begin{thebibliography}{59}%
\makeatletter
\providecommand \@ifxundefined [1]{%
 \@ifx{#1\undefined}
}%
\providecommand \@ifnum [1]{%
 \ifnum #1\expandafter \@firstoftwo
 \else \expandafter \@secondoftwo
 \fi
}%
\providecommand \@ifx [1]{%
 \ifx #1\expandafter \@firstoftwo
 \else \expandafter \@secondoftwo
 \fi
}%
\providecommand \natexlab [1]{#1}%
\providecommand \enquote  [1]{``#1''}%
\providecommand \bibnamefont  [1]{#1}%
\providecommand \bibfnamefont [1]{#1}%
\providecommand \citenamefont [1]{#1}%
\providecommand \href@noop [0]{\@secondoftwo}%
\providecommand \href [0]{\begingroup \@sanitize@url \@href}%
\providecommand \@href[1]{\@@startlink{#1}\@@href}%
\providecommand \@@href[1]{\endgroup#1\@@endlink}%
\providecommand \@sanitize@url [0]{\catcode `\\12\catcode `\$12\catcode
  `\&12\catcode `\#12\catcode `\^12\catcode `\_12\catcode `\%12\relax}%
\providecommand \@@startlink[1]{}%
\providecommand \@@endlink[0]{}%
\providecommand \url  [0]{\begingroup\@sanitize@url \@url }%
\providecommand \@url [1]{\endgroup\@href {#1}{\urlprefix }}%
\providecommand \urlprefix  [0]{URL }%
\providecommand \Eprint [0]{\href }%
\providecommand \doibase [0]{http://dx.doi.org/}%
\providecommand \selectlanguage [0]{\@gobble}%
\providecommand \bibinfo  [0]{\@secondoftwo}%
\providecommand \bibfield  [0]{\@secondoftwo}%
\providecommand \translation [1]{[#1]}%
\providecommand \BibitemOpen [0]{}%
\providecommand \bibitemStop [0]{}%
\providecommand \bibitemNoStop [0]{.\EOS\space}%
\providecommand \EOS [0]{\spacefactor3000\relax}%
\providecommand \BibitemShut  [1]{\csname bibitem#1\endcsname}%
\let\auto@bib@innerbib\@empty
\bibitem [{\citenamefont {Bertini}\ \emph {et~al.}(2016)\citenamefont
  {Bertini}, \citenamefont {Collura}, \citenamefont {{De Nardis}},\ and\
  \citenamefont {Fagotti}}]{PhysRevLett.117.207201}%
  \BibitemOpen
  \bibfield  {author} {\bibinfo {author} {\bibfnamefont {B.}~\bibnamefont
  {Bertini}}, \bibinfo {author} {\bibfnamefont {M.}~\bibnamefont {Collura}},
  \bibinfo {author} {\bibfnamefont {J.}~\bibnamefont {{De Nardis}}}, \ and\
  \bibinfo {author} {\bibfnamefont {M.}~\bibnamefont {Fagotti}},\ }\href
  {\doibase 10.1103/PhysRevLett.117.207201} {\bibfield  {journal} {\bibinfo
  {journal} {Phys. Rev. Lett.}\ }\textbf {\bibinfo {volume} {117}},\ \bibinfo
  {pages} {207201} (\bibinfo {year} {2016})}\BibitemShut {NoStop}%
\bibitem [{\citenamefont {Castro-Alvaredo}\ \emph {et~al.}(2016)\citenamefont
  {Castro-Alvaredo}, \citenamefont {Doyon},\ and\ \citenamefont
  {Yoshimura}}]{PhysRevX.6.041065}%
  \BibitemOpen
  \bibfield  {author} {\bibinfo {author} {\bibfnamefont {O.~A.}\ \bibnamefont
  {Castro-Alvaredo}}, \bibinfo {author} {\bibfnamefont {B.}~\bibnamefont
  {Doyon}}, \ and\ \bibinfo {author} {\bibfnamefont {T.}~\bibnamefont
  {Yoshimura}},\ }\href {\doibase 10.1103/PhysRevX.6.041065} {\bibfield
  {journal} {\bibinfo  {journal} {Phys. Rev. X}\ }\textbf {\bibinfo {volume}
  {6}},\ \bibinfo {pages} {041065} (\bibinfo {year} {2016})}\BibitemShut
  {NoStop}%
\bibitem [{\citenamefont {Doyon}\ \emph {et~al.}(2018)\citenamefont {Doyon},
  \citenamefont {Yoshimura},\ and\ \citenamefont
  {Caux}}]{PhysRevLett.120.045301}%
  \BibitemOpen
  \bibfield  {author} {\bibinfo {author} {\bibfnamefont {B.}~\bibnamefont
  {Doyon}}, \bibinfo {author} {\bibfnamefont {T.}~\bibnamefont {Yoshimura}}, \
  and\ \bibinfo {author} {\bibfnamefont {J.-S.}\ \bibnamefont {Caux}},\ }\href
  {\doibase 10.1103/PhysRevLett.120.045301} {\bibfield  {journal} {\bibinfo
  {journal} {Phys. Rev. Lett.}\ }\textbf {\bibinfo {volume} {120}},\ \bibinfo
  {pages} {045301} (\bibinfo {year} {2018})}\BibitemShut {NoStop}%
\bibitem [{\citenamefont {Piroli}\ \emph {et~al.}(2017)\citenamefont {Piroli},
  \citenamefont {{De Nardis}}, \citenamefont {Collura}, \citenamefont
  {Bertini},\ and\ \citenamefont {Fagotti}}]{PhysRevB.96.115124}%
  \BibitemOpen
  \bibfield  {author} {\bibinfo {author} {\bibfnamefont {L.}~\bibnamefont
  {Piroli}}, \bibinfo {author} {\bibfnamefont {J.}~\bibnamefont {{De Nardis}}},
  \bibinfo {author} {\bibfnamefont {M.}~\bibnamefont {Collura}}, \bibinfo
  {author} {\bibfnamefont {B.}~\bibnamefont {Bertini}}, \ and\ \bibinfo
  {author} {\bibfnamefont {M.}~\bibnamefont {Fagotti}},\ }\href {\doibase
  10.1103/PhysRevB.96.115124} {\bibfield  {journal} {\bibinfo  {journal} {Phys.
  Rev. B}\ }\textbf {\bibinfo {volume} {96}},\ \bibinfo {pages} {115124}
  (\bibinfo {year} {2017})}\BibitemShut {NoStop}%
\bibitem [{\citenamefont {Bulchandani}\ \emph {et~al.}(2017)\citenamefont
  {Bulchandani}, \citenamefont {Vasseur}, \citenamefont {Karrasch},\ and\
  \citenamefont {Moore}}]{Bulchandani2017}%
  \BibitemOpen
  \bibfield  {author} {\bibinfo {author} {\bibfnamefont {V.~B.}\ \bibnamefont
  {Bulchandani}}, \bibinfo {author} {\bibfnamefont {R.}~\bibnamefont
  {Vasseur}}, \bibinfo {author} {\bibfnamefont {C.}~\bibnamefont {Karrasch}}, \
  and\ \bibinfo {author} {\bibfnamefont {J.~E.}\ \bibnamefont {Moore}},\ }\href
  {\doibase 10.1103/physrevlett.119.220604} {\bibfield  {journal} {\bibinfo
  {journal} {Physical Review Letters}\ }\textbf {\bibinfo {volume} {119}}
  (\bibinfo {year} {2017}),\ 10.1103/physrevlett.119.220604}\BibitemShut
  {NoStop}%
\bibitem [{\citenamefont {Gopalakrishnan}\ \emph {et~al.}(2018)\citenamefont
  {Gopalakrishnan}, \citenamefont {Huse}, \citenamefont {Khemani},\ and\
  \citenamefont {Vasseur}}]{Gopalakrishnan2018}%
  \BibitemOpen
  \bibfield  {author} {\bibinfo {author} {\bibfnamefont {S.}~\bibnamefont
  {Gopalakrishnan}}, \bibinfo {author} {\bibfnamefont {D.~A.}\ \bibnamefont
  {Huse}}, \bibinfo {author} {\bibfnamefont {V.}~\bibnamefont {Khemani}}, \
  and\ \bibinfo {author} {\bibfnamefont {R.}~\bibnamefont {Vasseur}},\ }\href
  {\doibase 10.1103/physrevb.98.220303} {\bibfield  {journal} {\bibinfo
  {journal} {Physical Review B}\ }\textbf {\bibinfo {volume} {98}} (\bibinfo
  {year} {2018}),\ 10.1103/physrevb.98.220303}\BibitemShut {NoStop}%
\bibitem [{\citenamefont {Kovtun}\ \emph {et~al.}(2003)\citenamefont {Kovtun},
  \citenamefont {Son},\ and\ \citenamefont {Starinets}}]{Kovtun2003}%
  \BibitemOpen
  \bibfield  {author} {\bibinfo {author} {\bibfnamefont {P.}~\bibnamefont
  {Kovtun}}, \bibinfo {author} {\bibfnamefont {D.~T.}\ \bibnamefont {Son}}, \
  and\ \bibinfo {author} {\bibfnamefont {A.~O.}\ \bibnamefont {Starinets}},\
  }\href {\doibase 10.1088/1126-6708/2003/10/064} {\bibfield  {journal}
  {\bibinfo  {journal} {Journal of High Energy Physics}\ }\textbf {\bibinfo
  {volume} {2003}},\ \bibinfo {pages} {064} (\bibinfo {year}
  {2003})}\BibitemShut {NoStop}%
\bibitem [{\citenamefont {Grozdanov}\ \emph {et~al.}(2017)\citenamefont
  {Grozdanov}, \citenamefont {Hofman},\ and\ \citenamefont
  {Iqbal}}]{PhysRevD.95.096003}%
  \BibitemOpen
  \bibfield  {author} {\bibinfo {author} {\bibfnamefont {S.}~\bibnamefont
  {Grozdanov}}, \bibinfo {author} {\bibfnamefont {D.~M.}\ \bibnamefont
  {Hofman}}, \ and\ \bibinfo {author} {\bibfnamefont {N.}~\bibnamefont
  {Iqbal}},\ }\href {\doibase 10.1103/PhysRevD.95.096003} {\bibfield  {journal}
  {\bibinfo  {journal} {Phys. Rev. D}\ }\textbf {\bibinfo {volume} {95}},\
  \bibinfo {pages} {096003} (\bibinfo {year} {2017})}\BibitemShut {NoStop}%
\bibitem [{\citenamefont {Bhaseen}\ \emph {et~al.}(2015)\citenamefont
  {Bhaseen}, \citenamefont {Doyon}, \citenamefont {Lucas},\ and\ \citenamefont
  {Schalm}}]{Bhaseen:2015aa}%
  \BibitemOpen
  \bibfield  {author} {\bibinfo {author} {\bibfnamefont {M.~J.}\ \bibnamefont
  {Bhaseen}}, \bibinfo {author} {\bibfnamefont {B.}~\bibnamefont {Doyon}},
  \bibinfo {author} {\bibfnamefont {A.}~\bibnamefont {Lucas}}, \ and\ \bibinfo
  {author} {\bibfnamefont {K.}~\bibnamefont {Schalm}},\ }\href
  {http://dx.doi.org/10.1038/nphys3320} {\bibfield  {journal} {\bibinfo
  {journal} {Nat. Phys.}\ }\textbf {\bibinfo {volume} {11}},\ \bibinfo {pages}
  {509 EP } (\bibinfo {year} {2015})}\BibitemShut {NoStop}%
\bibitem [{\citenamefont {Sachdev}\ and\ \citenamefont
  {Young}(1997)}]{Sachdev19972}%
  \BibitemOpen
  \bibfield  {author} {\bibinfo {author} {\bibfnamefont {S.}~\bibnamefont
  {Sachdev}}\ and\ \bibinfo {author} {\bibfnamefont {A.~P.}\ \bibnamefont
  {Young}},\ }\href {\doibase 10.1103/physrevlett.78.2220} {\bibfield
  {journal} {\bibinfo  {journal} {Physical Review Letters}\ }\textbf {\bibinfo
  {volume} {78}},\ \bibinfo {pages} {2220} (\bibinfo {year}
  {1997})}\BibitemShut {NoStop}%
\bibitem [{\citenamefont {Damle}\ and\ \citenamefont
  {Sachdev}(1998)}]{Damle1998}%
  \BibitemOpen
  \bibfield  {author} {\bibinfo {author} {\bibfnamefont {K.}~\bibnamefont
  {Damle}}\ and\ \bibinfo {author} {\bibfnamefont {S.}~\bibnamefont
  {Sachdev}},\ }\href {\doibase 10.1103/physrevb.57.8307} {\bibfield  {journal}
  {\bibinfo  {journal} {Physical Review B}\ }\textbf {\bibinfo {volume} {57}},\
  \bibinfo {pages} {8307} (\bibinfo {year} {1998})}\BibitemShut {NoStop}%
\bibitem [{\citenamefont {Damle}\ and\ \citenamefont
  {Sachdev}(2005)}]{Damle2005}%
  \BibitemOpen
  \bibfield  {author} {\bibinfo {author} {\bibfnamefont {K.}~\bibnamefont
  {Damle}}\ and\ \bibinfo {author} {\bibfnamefont {S.}~\bibnamefont
  {Sachdev}},\ }\href {\doibase 10.1103/physrevlett.95.187201} {\bibfield
  {journal} {\bibinfo  {journal} {Physical Review Letters}\ }\textbf {\bibinfo
  {volume} {95}} (\bibinfo {year} {2005}),\
  10.1103/physrevlett.95.187201}\BibitemShut {NoStop}%
\bibitem [{\citenamefont {Haegeman}\ \emph {et~al.}(2016)\citenamefont
  {Haegeman}, \citenamefont {Lubich}, \citenamefont {Oseledets}, \citenamefont
  {Vandereycken},\ and\ \citenamefont {Verstraete}}]{Haegeman2016}%
  \BibitemOpen
  \bibfield  {author} {\bibinfo {author} {\bibfnamefont {J.}~\bibnamefont
  {Haegeman}}, \bibinfo {author} {\bibfnamefont {C.}~\bibnamefont {Lubich}},
  \bibinfo {author} {\bibfnamefont {I.}~\bibnamefont {Oseledets}}, \bibinfo
  {author} {\bibfnamefont {B.}~\bibnamefont {Vandereycken}}, \ and\ \bibinfo
  {author} {\bibfnamefont {F.}~\bibnamefont {Verstraete}},\ }\href {\doibase
  10.1103/physrevb.94.165116} {\bibfield  {journal} {\bibinfo  {journal}
  {Physical Review B}\ }\textbf {\bibinfo {volume} {94}} (\bibinfo {year}
  {2016}),\ 10.1103/physrevb.94.165116}\BibitemShut {NoStop}%
\bibitem [{\citenamefont {Haegeman}\ \emph {et~al.}(2013)\citenamefont
  {Haegeman}, \citenamefont {Osborne},\ and\ \citenamefont
  {Verstraete}}]{PhysRevB.88.075133}%
  \BibitemOpen
  \bibfield  {author} {\bibinfo {author} {\bibfnamefont {J.}~\bibnamefont
  {Haegeman}}, \bibinfo {author} {\bibfnamefont {T.~J.}\ \bibnamefont
  {Osborne}}, \ and\ \bibinfo {author} {\bibfnamefont {F.}~\bibnamefont
  {Verstraete}},\ }\href {\doibase 10.1103/PhysRevB.88.075133} {\bibfield
  {journal} {\bibinfo  {journal} {Phys. Rev. B}\ }\textbf {\bibinfo {volume}
  {88}},\ \bibinfo {pages} {075133} (\bibinfo {year} {2013})}\BibitemShut
  {NoStop}%
\bibitem [{\citenamefont {Hallam}\ \emph {et~al.}(2019)\citenamefont {Hallam},
  \citenamefont {Morley},\ and\ \citenamefont {Green}}]{Hallam2019}%
  \BibitemOpen
  \bibfield  {author} {\bibinfo {author} {\bibfnamefont {A.}~\bibnamefont
  {Hallam}}, \bibinfo {author} {\bibfnamefont {J.~G.}\ \bibnamefont {Morley}},
  \ and\ \bibinfo {author} {\bibfnamefont {A.~G.}\ \bibnamefont {Green}},\
  }\href {\doibase 10.1038/s41467-019-10336-4} {\bibfield  {journal} {\bibinfo
  {journal} {Nature Communications}\ }\textbf {\bibinfo {volume} {10}}
  (\bibinfo {year} {2019}),\ 10.1038/s41467-019-10336-4}\BibitemShut {NoStop}%
\bibitem [{\citenamefont {Leviatan}\ \emph {et~al.}(2017)\citenamefont
  {Leviatan}, \citenamefont {Pollmann}, \citenamefont {Bardarson},
  \citenamefont {Huse},\ and\ \citenamefont {Altman}}]{Leviatan2017}%
  \BibitemOpen
  \bibfield  {author} {\bibinfo {author} {\bibfnamefont {E.}~\bibnamefont
  {Leviatan}}, \bibinfo {author} {\bibfnamefont {F.}~\bibnamefont {Pollmann}},
  \bibinfo {author} {\bibfnamefont {J.~H.}\ \bibnamefont {Bardarson}}, \bibinfo
  {author} {\bibfnamefont {D.~A.}\ \bibnamefont {Huse}}, \ and\ \bibinfo
  {author} {\bibfnamefont {E.}~\bibnamefont {Altman}},\ }\href@noop {} {\
  (\bibinfo {year} {2017})},\ \Eprint {http://arxiv.org/abs/arXiv:1702.08894}
  {arXiv:1702.08894} \BibitemShut {NoStop}%
\bibitem [{\citenamefont {Paeckel}\ \emph {et~al.}(2019)\citenamefont
  {Paeckel}, \citenamefont {Köhler}, \citenamefont {Swoboda}, \citenamefont
  {Manmana}, \citenamefont {Schollwöck},\ and\ \citenamefont
  {Hubig}}]{Paeckel2019}%
  \BibitemOpen
  \bibfield  {author} {\bibinfo {author} {\bibfnamefont {S.}~\bibnamefont
  {Paeckel}}, \bibinfo {author} {\bibfnamefont {T.}~\bibnamefont {Köhler}},
  \bibinfo {author} {\bibfnamefont {A.}~\bibnamefont {Swoboda}}, \bibinfo
  {author} {\bibfnamefont {S.~R.}\ \bibnamefont {Manmana}}, \bibinfo {author}
  {\bibfnamefont {U.}~\bibnamefont {Schollwöck}}, \ and\ \bibinfo {author}
  {\bibfnamefont {C.}~\bibnamefont {Hubig}},\ }\href@noop {} {\enquote
  {\bibinfo {title} {Time-evolution methods for matrix-product states},}\ }
  (\bibinfo {year} {2019}),\ \Eprint {http://arxiv.org/abs/arXiv:1901.05824}
  {arXiv:1901.05824} \BibitemShut {NoStop}%
\bibitem [{\citenamefont {Kloss}\ \emph {et~al.}(2018)\citenamefont {Kloss},
  \citenamefont {Lev},\ and\ \citenamefont {Reichman}}]{TDVPProblems}%
  \BibitemOpen
  \bibfield  {author} {\bibinfo {author} {\bibfnamefont {B.}~\bibnamefont
  {Kloss}}, \bibinfo {author} {\bibfnamefont {Y.~B.}\ \bibnamefont {Lev}}, \
  and\ \bibinfo {author} {\bibfnamefont {D.}~\bibnamefont {Reichman}},\ }\href
  {\doibase 10.1103/PhysRevB.97.024307} {\bibfield  {journal} {\bibinfo
  {journal} {Phys. Rev. B}\ }\textbf {\bibinfo {volume} {97}},\ \bibinfo
  {pages} {024307} (\bibinfo {year} {2018})}\BibitemShut {NoStop}%
\bibitem [{\citenamefont {Nardis}\ \emph {et~al.}(2019)\citenamefont {Nardis},
  \citenamefont {Medenjak}, \citenamefont {Karrasch},\ and\ \citenamefont
  {Ilievski}}]{DeNardis2019Anomalous}%
  \BibitemOpen
  \bibfield  {author} {\bibinfo {author} {\bibfnamefont {J.~D.}\ \bibnamefont
  {Nardis}}, \bibinfo {author} {\bibfnamefont {M.}~\bibnamefont {Medenjak}},
  \bibinfo {author} {\bibfnamefont {C.}~\bibnamefont {Karrasch}}, \ and\
  \bibinfo {author} {\bibfnamefont {E.}~\bibnamefont {Ilievski}},\ }\href@noop
  {} {\  (\bibinfo {year} {2019})},\ \Eprint
  {http://arxiv.org/abs/arXiv:1903.07598} {arXiv:1903.07598} \BibitemShut
  {NoStop}%
\bibitem [{\citenamefont {Richter}\ \emph {et~al.}(2019)\citenamefont
  {Richter}, \citenamefont {Casper}, \citenamefont {Brenig},\ and\
  \citenamefont {Steinigeweg}}]{Richter2019}%
  \BibitemOpen
  \bibfield  {author} {\bibinfo {author} {\bibfnamefont {J.}~\bibnamefont
  {Richter}}, \bibinfo {author} {\bibfnamefont {N.}~\bibnamefont {Casper}},
  \bibinfo {author} {\bibfnamefont {W.}~\bibnamefont {Brenig}}, \ and\ \bibinfo
  {author} {\bibfnamefont {R.}~\bibnamefont {Steinigeweg}},\ }\href@noop {} {\
  (\bibinfo {year} {2019})},\ \Eprint {http://arxiv.org/abs/arXiv:1907.03004}
  {arXiv:1907.03004} \BibitemShut {NoStop}%
\bibitem [{\citenamefont {Dupont}\ \emph {et~al.}(2018)\citenamefont {Dupont},
  \citenamefont {Capponi}, \citenamefont {Laflorencie},\ and\ \citenamefont
  {Orignac}}]{Dupont2018}%
  \BibitemOpen
  \bibfield  {author} {\bibinfo {author} {\bibfnamefont {M.}~\bibnamefont
  {Dupont}}, \bibinfo {author} {\bibfnamefont {S.}~\bibnamefont {Capponi}},
  \bibinfo {author} {\bibfnamefont {N.}~\bibnamefont {Laflorencie}}, \ and\
  \bibinfo {author} {\bibfnamefont {E.}~\bibnamefont {Orignac}},\ }\href
  {\doibase 10.1103/physrevb.98.094403} {\bibfield  {journal} {\bibinfo
  {journal} {Physical Review B}\ }\textbf {\bibinfo {volume} {98}} (\bibinfo
  {year} {2018}),\ 10.1103/physrevb.98.094403}\BibitemShut {NoStop}%
\bibitem [{\citenamefont {Dupont}\ and\ \citenamefont
  {Moore}(2019)}]{DupontMoore2019}%
  \BibitemOpen
  \bibfield  {author} {\bibinfo {author} {\bibfnamefont {M.}~\bibnamefont
  {Dupont}}\ and\ \bibinfo {author} {\bibfnamefont {J.~E.}\ \bibnamefont
  {Moore}},\ }\href@noop {} {\  (\bibinfo {year} {2019})},\ \Eprint
  {http://arxiv.org/abs/arXiv:1907.12115} {arXiv:1907.12115} \BibitemShut
  {NoStop}%
\bibitem [{\citenamefont {Antal}\ \emph {et~al.}(1999)\citenamefont {Antal},
  \citenamefont {R\'acz}, \citenamefont {R\'akos},\ and\ \citenamefont
  {Sch\"utz}}]{PhysRevE.59.4912}%
  \BibitemOpen
  \bibfield  {author} {\bibinfo {author} {\bibfnamefont {T.}~\bibnamefont
  {Antal}}, \bibinfo {author} {\bibfnamefont {Z.}~\bibnamefont {R\'acz}},
  \bibinfo {author} {\bibfnamefont {A.}~\bibnamefont {R\'akos}}, \ and\
  \bibinfo {author} {\bibfnamefont {G.~M.}\ \bibnamefont {Sch\"utz}},\ }\href
  {\doibase 10.1103/PhysRevE.59.4912} {\bibfield  {journal} {\bibinfo
  {journal} {Phys. Rev. E}\ }\textbf {\bibinfo {volume} {59}},\ \bibinfo
  {pages} {4912} (\bibinfo {year} {1999})}\BibitemShut {NoStop}%
\bibitem [{\citenamefont {Gobert}\ \emph {et~al.}(2005)\citenamefont {Gobert},
  \citenamefont {Kollath}, \citenamefont {Schollw\"ock},\ and\ \citenamefont
  {Sch\"utz}}]{PhysRevE.71.036102}%
  \BibitemOpen
  \bibfield  {author} {\bibinfo {author} {\bibfnamefont {D.}~\bibnamefont
  {Gobert}}, \bibinfo {author} {\bibfnamefont {C.}~\bibnamefont {Kollath}},
  \bibinfo {author} {\bibfnamefont {U.}~\bibnamefont {Schollw\"ock}}, \ and\
  \bibinfo {author} {\bibfnamefont {G.}~\bibnamefont {Sch\"utz}},\ }\href
  {\doibase 10.1103/PhysRevE.71.036102} {\bibfield  {journal} {\bibinfo
  {journal} {Phys. Rev. E}\ }\textbf {\bibinfo {volume} {71}},\ \bibinfo
  {pages} {036102} (\bibinfo {year} {2005})}\BibitemShut {NoStop}%
\bibitem [{\citenamefont {Santos}(2008)}]{PhysRevE.78.031125}%
  \BibitemOpen
  \bibfield  {author} {\bibinfo {author} {\bibfnamefont {L.~F.}\ \bibnamefont
  {Santos}},\ }\href {\doibase 10.1103/PhysRevE.78.031125} {\bibfield
  {journal} {\bibinfo  {journal} {Phys. Rev. E}\ }\textbf {\bibinfo {volume}
  {78}},\ \bibinfo {pages} {031125} (\bibinfo {year} {2008})}\BibitemShut
  {NoStop}%
\bibitem [{\citenamefont {Antal}\ \emph {et~al.}(2008)\citenamefont {Antal},
  \citenamefont {Krapivsky},\ and\ \citenamefont
  {R\'akos}}]{PhysRevE.78.061115}%
  \BibitemOpen
  \bibfield  {author} {\bibinfo {author} {\bibfnamefont {T.}~\bibnamefont
  {Antal}}, \bibinfo {author} {\bibfnamefont {P.~L.}\ \bibnamefont
  {Krapivsky}}, \ and\ \bibinfo {author} {\bibfnamefont {A.}~\bibnamefont
  {R\'akos}},\ }\href {\doibase 10.1103/PhysRevE.78.061115} {\bibfield
  {journal} {\bibinfo  {journal} {Phys. Rev. E}\ }\textbf {\bibinfo {volume}
  {78}},\ \bibinfo {pages} {061115} (\bibinfo {year} {2008})}\BibitemShut
  {NoStop}%
\bibitem [{\citenamefont {Santos}(2009)}]{santos2009transport}%
  \BibitemOpen
  \bibfield  {author} {\bibinfo {author} {\bibfnamefont {L.~F.}\ \bibnamefont
  {Santos}},\ }\href@noop {} {\bibfield  {journal} {\bibinfo  {journal}
  {Journal of Mathematical Physics}\ }\textbf {\bibinfo {volume} {50}},\
  \bibinfo {pages} {095211} (\bibinfo {year} {2009})}\BibitemShut {NoStop}%
\bibitem [{\citenamefont {Lancaster}\ and\ \citenamefont
  {Mitra}(2010)}]{lancaster2010quantum}%
  \BibitemOpen
  \bibfield  {author} {\bibinfo {author} {\bibfnamefont {J.}~\bibnamefont
  {Lancaster}}\ and\ \bibinfo {author} {\bibfnamefont {A.}~\bibnamefont
  {Mitra}},\ }\href@noop {} {\bibfield  {journal} {\bibinfo  {journal}
  {Physical Review E}\ }\textbf {\bibinfo {volume} {81}},\ \bibinfo {pages}
  {061134} (\bibinfo {year} {2010})}\BibitemShut {NoStop}%
\bibitem [{\citenamefont {Santos}\ and\ \citenamefont
  {Mitra}(2011)}]{PhysRevE.84.016206}%
  \BibitemOpen
  \bibfield  {author} {\bibinfo {author} {\bibfnamefont {L.~F.}\ \bibnamefont
  {Santos}}\ and\ \bibinfo {author} {\bibfnamefont {A.}~\bibnamefont {Mitra}},\
  }\href {\doibase 10.1103/PhysRevE.84.016206} {\bibfield  {journal} {\bibinfo
  {journal} {Phys. Rev. E}\ }\textbf {\bibinfo {volume} {84}},\ \bibinfo
  {pages} {016206} (\bibinfo {year} {2011})}\BibitemShut {NoStop}%
\bibitem [{\citenamefont {Sabetta}\ and\ \citenamefont
  {Misguich}(2013)}]{PhysRevB.88.245114}%
  \BibitemOpen
  \bibfield  {author} {\bibinfo {author} {\bibfnamefont {T.}~\bibnamefont
  {Sabetta}}\ and\ \bibinfo {author} {\bibfnamefont {G.}~\bibnamefont
  {Misguich}},\ }\href {\doibase 10.1103/PhysRevB.88.245114} {\bibfield
  {journal} {\bibinfo  {journal} {Phys. Rev. B}\ }\textbf {\bibinfo {volume}
  {88}},\ \bibinfo {pages} {245114} (\bibinfo {year} {2013})}\BibitemShut
  {NoStop}%
\bibitem [{\citenamefont {Gamayun}\ \emph {et~al.}(2019)\citenamefont
  {Gamayun}, \citenamefont {Miao},\ and\ \citenamefont
  {Ilievski}}]{Gamayun2019}%
  \BibitemOpen
  \bibfield  {author} {\bibinfo {author} {\bibfnamefont {O.}~\bibnamefont
  {Gamayun}}, \bibinfo {author} {\bibfnamefont {Y.}~\bibnamefont {Miao}}, \
  and\ \bibinfo {author} {\bibfnamefont {E.}~\bibnamefont {Ilievski}},\ }\href
  {https://arxiv.org/abs/1901.08944} {\bibfield  {journal} {\bibinfo  {journal}
  {arXiv:1901.08944}\ } (\bibinfo {year} {2019})}\BibitemShut {NoStop}%
\bibitem [{\citenamefont {Misguich}\ \emph {et~al.}(2019)\citenamefont
  {Misguich}, \citenamefont {Pavloff},\ and\ \citenamefont
  {Pasquier}}]{Misguich2019}%
  \BibitemOpen
  \bibfield  {author} {\bibinfo {author} {\bibfnamefont {G.}~\bibnamefont
  {Misguich}}, \bibinfo {author} {\bibfnamefont {N.}~\bibnamefont {Pavloff}}, \
  and\ \bibinfo {author} {\bibfnamefont {V.}~\bibnamefont {Pasquier}},\ }\href
  {\doibase 10.21468/SciPostPhys.7.2.025} {\bibfield  {journal} {\bibinfo
  {journal} {SciPost Phys.}\ }\textbf {\bibinfo {volume} {7}},\ \bibinfo
  {pages} {25} (\bibinfo {year} {2019})}\BibitemShut {NoStop}%
\bibitem [{\citenamefont {Bulchandani}\ and\ \citenamefont
  {Karrasch}(2019)}]{PhysRevB.99.121410}%
  \BibitemOpen
  \bibfield  {author} {\bibinfo {author} {\bibfnamefont {V.~B.}\ \bibnamefont
  {Bulchandani}}\ and\ \bibinfo {author} {\bibfnamefont {C.}~\bibnamefont
  {Karrasch}},\ }\href {\doibase 10.1103/PhysRevB.99.121410} {\bibfield
  {journal} {\bibinfo  {journal} {Phys. Rev. B}\ }\textbf {\bibinfo {volume}
  {99}},\ \bibinfo {pages} {121410} (\bibinfo {year} {2019})}\BibitemShut
  {NoStop}%
\bibitem [{\citenamefont {y.~Choi}\ \emph {et~al.}(2016)\citenamefont
  {y.~Choi}, \citenamefont {Hild}, \citenamefont {Zeiher}, \citenamefont
  {Schauss}, \citenamefont {Rubio-Abadal}, \citenamefont {Yefsah},
  \citenamefont {Khemani}, \citenamefont {Huse}, \citenamefont {Bloch},\ and\
  \citenamefont {Gross}}]{Choi2016}%
  \BibitemOpen
  \bibfield  {author} {\bibinfo {author} {\bibfnamefont {J.}~\bibnamefont
  {y.~Choi}}, \bibinfo {author} {\bibfnamefont {S.}~\bibnamefont {Hild}},
  \bibinfo {author} {\bibfnamefont {J.}~\bibnamefont {Zeiher}}, \bibinfo
  {author} {\bibfnamefont {P.}~\bibnamefont {Schauss}}, \bibinfo {author}
  {\bibfnamefont {A.}~\bibnamefont {Rubio-Abadal}}, \bibinfo {author}
  {\bibfnamefont {T.}~\bibnamefont {Yefsah}}, \bibinfo {author} {\bibfnamefont
  {V.}~\bibnamefont {Khemani}}, \bibinfo {author} {\bibfnamefont {D.~A.}\
  \bibnamefont {Huse}}, \bibinfo {author} {\bibfnamefont {I.}~\bibnamefont
  {Bloch}}, \ and\ \bibinfo {author} {\bibfnamefont {C.}~\bibnamefont
  {Gross}},\ }\href {\doibase 10.1126/science.aaf8834} {\bibfield  {journal}
  {\bibinfo  {journal} {Science}\ }\textbf {\bibinfo {volume} {352}},\ \bibinfo
  {pages} {1547} (\bibinfo {year} {2016})}\BibitemShut {NoStop}%
\bibitem [{\citenamefont {Vidmar}\ \emph {et~al.}(2015)\citenamefont {Vidmar},
  \citenamefont {Ronzheimer}, \citenamefont {Schreiber}, \citenamefont {Braun},
  \citenamefont {Hodgman}, \citenamefont {Langer}, \citenamefont
  {Heidrich-Meisner}, \citenamefont {Bloch},\ and\ \citenamefont
  {Schneider}}]{PhysRevLett.115.175301}%
  \BibitemOpen
  \bibfield  {author} {\bibinfo {author} {\bibfnamefont {L.}~\bibnamefont
  {Vidmar}}, \bibinfo {author} {\bibfnamefont {J.~P.}\ \bibnamefont
  {Ronzheimer}}, \bibinfo {author} {\bibfnamefont {M.}~\bibnamefont
  {Schreiber}}, \bibinfo {author} {\bibfnamefont {S.}~\bibnamefont {Braun}},
  \bibinfo {author} {\bibfnamefont {S.~S.}\ \bibnamefont {Hodgman}}, \bibinfo
  {author} {\bibfnamefont {S.}~\bibnamefont {Langer}}, \bibinfo {author}
  {\bibfnamefont {F.}~\bibnamefont {Heidrich-Meisner}}, \bibinfo {author}
  {\bibfnamefont {I.}~\bibnamefont {Bloch}}, \ and\ \bibinfo {author}
  {\bibfnamefont {U.}~\bibnamefont {Schneider}},\ }\href {\doibase
  10.1103/PhysRevLett.115.175301} {\bibfield  {journal} {\bibinfo  {journal}
  {Phys. Rev. Lett.}\ }\textbf {\bibinfo {volume} {115}},\ \bibinfo {pages}
  {175301} (\bibinfo {year} {2015})}\BibitemShut {NoStop}%
\bibitem [{\citenamefont {Hauschild}\ \emph {et~al.}(2015)\citenamefont
  {Hauschild}, \citenamefont {Pollmann},\ and\ \citenamefont
  {Heidrich-Meisner}}]{PhysRevA.92.053629}%
  \BibitemOpen
  \bibfield  {author} {\bibinfo {author} {\bibfnamefont {J.}~\bibnamefont
  {Hauschild}}, \bibinfo {author} {\bibfnamefont {F.}~\bibnamefont {Pollmann}},
  \ and\ \bibinfo {author} {\bibfnamefont {F.}~\bibnamefont
  {Heidrich-Meisner}},\ }\href {\doibase 10.1103/PhysRevA.92.053629} {\bibfield
   {journal} {\bibinfo  {journal} {Phys. Rev. A}\ }\textbf {\bibinfo {volume}
  {92}},\ \bibinfo {pages} {053629} (\bibinfo {year} {2015})}\BibitemShut
  {NoStop}%
\bibitem [{\citenamefont {Smith}\ \emph {et~al.}(2019)\citenamefont {Smith},
  \citenamefont {Kim}, \citenamefont {Pollmann},\ and\ \citenamefont
  {Knolle}}]{1906.06343}%
  \BibitemOpen
  \bibfield  {author} {\bibinfo {author} {\bibfnamefont {A.}~\bibnamefont
  {Smith}}, \bibinfo {author} {\bibfnamefont {M.~S.}\ \bibnamefont {Kim}},
  \bibinfo {author} {\bibfnamefont {F.}~\bibnamefont {Pollmann}}, \ and\
  \bibinfo {author} {\bibfnamefont {J.}~\bibnamefont {Knolle}},\ }\href@noop {}
  {\  (\bibinfo {year} {2019})},\ \Eprint
  {http://arxiv.org/abs/arXiv:1906.06343} {arXiv:1906.06343} \BibitemShut
  {NoStop}%
\bibitem [{\citenamefont {Haegeman}\ \emph {et~al.}(2017)\citenamefont
  {Haegeman}, \citenamefont {Draxler}, \citenamefont {Stojevic}, \citenamefont
  {Cirac}, \citenamefont {Osborne},\ and\ \citenamefont
  {Verstraete}}]{SciPostPhys.3.1.006}%
  \BibitemOpen
  \bibfield  {author} {\bibinfo {author} {\bibfnamefont {J.}~\bibnamefont
  {Haegeman}}, \bibinfo {author} {\bibfnamefont {D.}~\bibnamefont {Draxler}},
  \bibinfo {author} {\bibfnamefont {V.}~\bibnamefont {Stojevic}}, \bibinfo
  {author} {\bibfnamefont {J.~I.}\ \bibnamefont {Cirac}}, \bibinfo {author}
  {\bibfnamefont {T.~J.}\ \bibnamefont {Osborne}}, \ and\ \bibinfo {author}
  {\bibfnamefont {F.}~\bibnamefont {Verstraete}},\ }\href {\doibase
  10.21468/SciPostPhys.3.1.006} {\bibfield  {journal} {\bibinfo  {journal}
  {SciPost Phys.}\ }\textbf {\bibinfo {volume} {3}},\ \bibinfo {pages} {006}
  (\bibinfo {year} {2017})}\BibitemShut {NoStop}%
\bibitem [{\citenamefont {Faddeev}\ and\ \citenamefont
  {Takhtajan}(1987)}]{Faddeev1987}%
  \BibitemOpen
  \bibfield  {author} {\bibinfo {author} {\bibfnamefont {L.~D.}\ \bibnamefont
  {Faddeev}}\ and\ \bibinfo {author} {\bibfnamefont {L.~A.}\ \bibnamefont
  {Takhtajan}},\ }\href {\doibase 10.1007/978-3-540-69969-9} {\emph {\bibinfo
  {title} {Hamiltonian Methods in the Theory of Solitons}}}\ (\bibinfo
  {publisher} {Springer Berlin Heidelberg},\ \bibinfo {year}
  {1987})\BibitemShut {NoStop}%
\bibitem [{\citenamefont {Takhtajan}(1977)}]{Takhtajan1977}%
  \BibitemOpen
  \bibfield  {author} {\bibinfo {author} {\bibfnamefont {L.}~\bibnamefont
  {Takhtajan}},\ }\href {\doibase 10.1016/0375-9601(77)90727-7} {\bibfield
  {journal} {\bibinfo  {journal} {Physics Letters A}\ }\textbf {\bibinfo
  {volume} {64}},\ \bibinfo {pages} {235} (\bibinfo {year} {1977})}\BibitemShut
  {NoStop}%
\bibitem [{\citenamefont {Calabrese}\ \emph {et~al.}(2008)\citenamefont
  {Calabrese}, \citenamefont {Hagendorf},\ and\ \citenamefont
  {Doussal}}]{Calabrese2008}%
  \BibitemOpen
  \bibfield  {author} {\bibinfo {author} {\bibfnamefont {P.}~\bibnamefont
  {Calabrese}}, \bibinfo {author} {\bibfnamefont {C.}~\bibnamefont
  {Hagendorf}}, \ and\ \bibinfo {author} {\bibfnamefont {P.~L.}\ \bibnamefont
  {Doussal}},\ }\href {\doibase 10.1088/1742-5468/2008/07/p07013} {\bibfield
  {journal} {\bibinfo  {journal} {Journal of Statistical Mechanics: Theory and
  Experiment}\ }\textbf {\bibinfo {volume} {2008}},\ \bibinfo {pages} {P07013}
  (\bibinfo {year} {2008})}\BibitemShut {NoStop}%
\bibitem [{\citenamefont {Mossel}\ and\ \citenamefont
  {Caux}(2010)}]{Mossel2010}%
  \BibitemOpen
  \bibfield  {author} {\bibinfo {author} {\bibfnamefont {J.}~\bibnamefont
  {Mossel}}\ and\ \bibinfo {author} {\bibfnamefont {J.-S.}\ \bibnamefont
  {Caux}},\ }\href {\doibase 10.1088/1367-2630/12/5/055028} {\bibfield
  {journal} {\bibinfo  {journal} {New Journal of Physics}\ }\textbf {\bibinfo
  {volume} {12}},\ \bibinfo {pages} {055028} (\bibinfo {year}
  {2010})}\BibitemShut {NoStop}%
\bibitem [{\citenamefont {Ljubotina}\ \emph
  {et~al.}(2017{\natexlab{a}})\citenamefont {Ljubotina}, \citenamefont
  {{\v{Z}}nidari{\v{c}}},\ and\ \citenamefont {Prosen}}]{Ljubotina2017}%
  \BibitemOpen
  \bibfield  {author} {\bibinfo {author} {\bibfnamefont {M.}~\bibnamefont
  {Ljubotina}}, \bibinfo {author} {\bibfnamefont {M.}~\bibnamefont
  {{\v{Z}}nidari{\v{c}}}}, \ and\ \bibinfo {author} {\bibfnamefont
  {T.}~\bibnamefont {Prosen}},\ }\href {\doibase 10.1038/ncomms16117}
  {\bibfield  {journal} {\bibinfo  {journal} {Nature Communications}\ }\textbf
  {\bibinfo {volume} {8}} (\bibinfo {year} {2017}{\natexlab{a}}),\
  10.1038/ncomms16117}\BibitemShut {NoStop}%
\bibitem [{\citenamefont {Misguich}\ \emph
  {et~al.}(2017{\natexlab{a}})\citenamefont {Misguich}, \citenamefont
  {Mallick},\ and\ \citenamefont {Krapivsky}}]{Misguich2017}%
  \BibitemOpen
  \bibfield  {author} {\bibinfo {author} {\bibfnamefont {G.}~\bibnamefont
  {Misguich}}, \bibinfo {author} {\bibfnamefont {K.}~\bibnamefont {Mallick}}, \
  and\ \bibinfo {author} {\bibfnamefont {P.~L.}\ \bibnamefont {Krapivsky}},\
  }\href {\doibase 10.1103/PhysRevB.96.195151} {\bibfield  {journal} {\bibinfo
  {journal} {Phys. Rev. B}\ }\textbf {\bibinfo {volume} {96}},\ \bibinfo
  {pages} {195151} (\bibinfo {year} {2017}{\natexlab{a}})}\BibitemShut
  {NoStop}%
\bibitem [{\citenamefont {St{\'{e}}phan}(2017)}]{Stephan17}%
  \BibitemOpen
  \bibfield  {author} {\bibinfo {author} {\bibfnamefont {J.-M.}\ \bibnamefont
  {St{\'{e}}phan}},\ }\href {\doibase 10.1088/1742-5468/aa8c19} {\bibfield
  {journal} {\bibinfo  {journal} {Journal of Statistical Mechanics: Theory and
  Experiment}\ }\textbf {\bibinfo {volume} {2017}},\ \bibinfo {pages} {103108}
  (\bibinfo {year} {2017})}\BibitemShut {NoStop}%
\bibitem [{\citenamefont {Ljubotina}\ \emph
  {et~al.}(2017{\natexlab{b}})\citenamefont {Ljubotina}, \citenamefont
  {{\v{Z}}nidari{\v{c}}},\ and\ \citenamefont {Prosen}}]{Ljubotina2017-diff}%
  \BibitemOpen
  \bibfield  {author} {\bibinfo {author} {\bibfnamefont {M.}~\bibnamefont
  {Ljubotina}}, \bibinfo {author} {\bibfnamefont {M.}~\bibnamefont
  {{\v{Z}}nidari{\v{c}}}}, \ and\ \bibinfo {author} {\bibfnamefont
  {T.}~\bibnamefont {Prosen}},\ }\href {\doibase 10.1088/1751-8121/aa8bdc}
  {\bibfield  {journal} {\bibinfo  {journal} {Journal of Physics A:
  Mathematical and Theoretical}\ }\textbf {\bibinfo {volume} {50}},\ \bibinfo
  {pages} {475002} (\bibinfo {year} {2017}{\natexlab{b}})}\BibitemShut
  {NoStop}%
\bibitem [{\citenamefont {Collura}\ \emph {et~al.}(2018)\citenamefont
  {Collura}, \citenamefont {Luca},\ and\ \citenamefont {Viti}}]{CDV18}%
  \BibitemOpen
  \bibfield  {author} {\bibinfo {author} {\bibfnamefont {M.}~\bibnamefont
  {Collura}}, \bibinfo {author} {\bibfnamefont {A.~D.}\ \bibnamefont {Luca}}, \
  and\ \bibinfo {author} {\bibfnamefont {J.}~\bibnamefont {Viti}},\ }\href
  {\doibase 10.1103/physrevb.97.081111} {\bibfield  {journal} {\bibinfo
  {journal} {Physical Review B}\ }\textbf {\bibinfo {volume} {97}} (\bibinfo
  {year} {2018}),\ 10.1103/physrevb.97.081111}\BibitemShut {NoStop}%
\bibitem [{\citenamefont {Prosen}(2011)}]{Prosen11}%
  \BibitemOpen
  \bibfield  {author} {\bibinfo {author} {\bibfnamefont {T.}~\bibnamefont
  {Prosen}},\ }\href {\doibase 10.1103/physrevlett.106.217206} {\bibfield
  {journal} {\bibinfo  {journal} {Physical Review Letters}\ }\textbf {\bibinfo
  {volume} {106}} (\bibinfo {year} {2011}),\
  10.1103/physrevlett.106.217206}\BibitemShut {NoStop}%
\bibitem [{\citenamefont {Ilievski}\ \emph {et~al.}(2016)\citenamefont
  {Ilievski}, \citenamefont {Medenjak}, \citenamefont {Prosen},\ and\
  \citenamefont {Zadnik}}]{ilievski2016quasilocal}%
  \BibitemOpen
  \bibfield  {author} {\bibinfo {author} {\bibfnamefont {E.}~\bibnamefont
  {Ilievski}}, \bibinfo {author} {\bibfnamefont {M.}~\bibnamefont {Medenjak}},
  \bibinfo {author} {\bibfnamefont {T.}~\bibnamefont {Prosen}}, \ and\ \bibinfo
  {author} {\bibfnamefont {L.}~\bibnamefont {Zadnik}},\ }\href {\doibase
  10.1088/1742-5468/2016/06/064008} {\bibfield  {journal} {\bibinfo  {journal}
  {Journal of Statistical Mechanics: Theory and Experiment}\ }\textbf {\bibinfo
  {volume} {2016}},\ \bibinfo {pages} {064008} (\bibinfo {year}
  {2016})}\BibitemShut {NoStop}%
\bibitem [{\citenamefont {Prosen}\ and\ \citenamefont
  {Ilievski}(2013)}]{PhysRevLett.111.057203}%
  \BibitemOpen
  \bibfield  {author} {\bibinfo {author} {\bibfnamefont {T.}~\bibnamefont
  {Prosen}}\ and\ \bibinfo {author} {\bibfnamefont {E.}~\bibnamefont
  {Ilievski}},\ }\href {\doibase 10.1103/PhysRevLett.111.057203} {\bibfield
  {journal} {\bibinfo  {journal} {Phys. Rev. Lett.}\ }\textbf {\bibinfo
  {volume} {111}},\ \bibinfo {pages} {057203} (\bibinfo {year}
  {2013})}\BibitemShut {NoStop}%
\bibitem [{\citenamefont {De~Luca}\ \emph {et~al.}(2017)\citenamefont
  {De~Luca}, \citenamefont {Collura},\ and\ \citenamefont {{De
  Nardis}}}]{PhysRevB.96.020403}%
  \BibitemOpen
  \bibfield  {author} {\bibinfo {author} {\bibfnamefont {A.}~\bibnamefont
  {De~Luca}}, \bibinfo {author} {\bibfnamefont {M.}~\bibnamefont {Collura}}, \
  and\ \bibinfo {author} {\bibfnamefont {J.}~\bibnamefont {{De Nardis}}},\
  }\href {\doibase 10.1103/PhysRevB.96.020403} {\bibfield  {journal} {\bibinfo
  {journal} {Phys. Rev. B}\ }\textbf {\bibinfo {volume} {96}},\ \bibinfo
  {pages} {020403} (\bibinfo {year} {2017})}\BibitemShut {NoStop}%
\bibitem [{\citenamefont {Vidal}(2004)}]{PhysRevLett.93.040502}%
  \BibitemOpen
  \bibfield  {author} {\bibinfo {author} {\bibfnamefont {G.}~\bibnamefont
  {Vidal}},\ }\href {\doibase 10.1103/PhysRevLett.93.040502} {\bibfield
  {journal} {\bibinfo  {journal} {Phys. Rev. Lett.}\ }\textbf {\bibinfo
  {volume} {93}},\ \bibinfo {pages} {040502} (\bibinfo {year}
  {2004})}\BibitemShut {NoStop}%
\bibitem [{\citenamefont {Haegeman}\ \emph {et~al.}(2011)\citenamefont
  {Haegeman}, \citenamefont {Cirac}, \citenamefont {Osborne}, \citenamefont
  {Pi{\v{z}}orn}, \citenamefont {Verschelde},\ and\ \citenamefont
  {Verstraete}}]{Haegeman2011}%
  \BibitemOpen
  \bibfield  {author} {\bibinfo {author} {\bibfnamefont {J.}~\bibnamefont
  {Haegeman}}, \bibinfo {author} {\bibfnamefont {J.~I.}\ \bibnamefont {Cirac}},
  \bibinfo {author} {\bibfnamefont {T.~J.}\ \bibnamefont {Osborne}}, \bibinfo
  {author} {\bibfnamefont {I.}~\bibnamefont {Pi{\v{z}}orn}}, \bibinfo {author}
  {\bibfnamefont {H.}~\bibnamefont {Verschelde}}, \ and\ \bibinfo {author}
  {\bibfnamefont {F.}~\bibnamefont {Verstraete}},\ }\href {\doibase
  10.1103/physrevlett.107.070601} {\bibfield  {journal} {\bibinfo  {journal}
  {Physical Review Letters}\ }\textbf {\bibinfo {volume} {107}} (\bibinfo
  {year} {2011}),\ 10.1103/physrevlett.107.070601}\BibitemShut {NoStop}%
\bibitem [{\citenamefont {Misguich}\ \emph
  {et~al.}(2017{\natexlab{b}})\citenamefont {Misguich}, \citenamefont
  {Mallick},\ and\ \citenamefont {Krapivsky}}]{Misguich17}%
  \BibitemOpen
  \bibfield  {author} {\bibinfo {author} {\bibfnamefont {G.}~\bibnamefont
  {Misguich}}, \bibinfo {author} {\bibfnamefont {K.}~\bibnamefont {Mallick}}, \
  and\ \bibinfo {author} {\bibfnamefont {P.~L.}\ \bibnamefont {Krapivsky}},\
  }\href {\doibase 10.1103/physrevb.96.195151} {\bibfield  {journal} {\bibinfo
  {journal} {Physical Review B}\ }\textbf {\bibinfo {volume} {96}} (\bibinfo
  {year} {2017}{\natexlab{b}}),\ 10.1103/physrevb.96.195151}\BibitemShut
  {NoStop}%
\bibitem [{\citenamefont {Choi}\ \emph {et~al.}(2019)\citenamefont {Choi},
  \citenamefont {Turner}, \citenamefont {Pichler}, \citenamefont {Ho},
  \citenamefont {Michailidis}, \citenamefont {Papi\ifmmode~\acute{c}\else
  \'{c}\fi{}}, \citenamefont {Serbyn}, \citenamefont {Lukin},\ and\
  \citenamefont {Abanin}}]{PhysRevLett.122.220603}%
  \BibitemOpen
  \bibfield  {author} {\bibinfo {author} {\bibfnamefont {S.}~\bibnamefont
  {Choi}}, \bibinfo {author} {\bibfnamefont {C.~J.}\ \bibnamefont {Turner}},
  \bibinfo {author} {\bibfnamefont {H.}~\bibnamefont {Pichler}}, \bibinfo
  {author} {\bibfnamefont {W.~W.}\ \bibnamefont {Ho}}, \bibinfo {author}
  {\bibfnamefont {A.~A.}\ \bibnamefont {Michailidis}}, \bibinfo {author}
  {\bibfnamefont {Z.}~\bibnamefont {Papi\ifmmode~\acute{c}\else \'{c}\fi{}}},
  \bibinfo {author} {\bibfnamefont {M.}~\bibnamefont {Serbyn}}, \bibinfo
  {author} {\bibfnamefont {M.~D.}\ \bibnamefont {Lukin}}, \ and\ \bibinfo
  {author} {\bibfnamefont {D.~A.}\ \bibnamefont {Abanin}},\ }\href {\doibase
  10.1103/PhysRevLett.122.220603} {\bibfield  {journal} {\bibinfo  {journal}
  {Phys. Rev. Lett.}\ }\textbf {\bibinfo {volume} {122}},\ \bibinfo {pages}
  {220603} (\bibinfo {year} {2019})}\BibitemShut {NoStop}%
\bibitem [{\citenamefont {Ho}\ \emph {et~al.}(2019)\citenamefont {Ho},
  \citenamefont {Choi}, \citenamefont {Pichler},\ and\ \citenamefont
  {Lukin}}]{PhysRevLett.122.040603}%
  \BibitemOpen
  \bibfield  {author} {\bibinfo {author} {\bibfnamefont {W.~W.}\ \bibnamefont
  {Ho}}, \bibinfo {author} {\bibfnamefont {S.}~\bibnamefont {Choi}}, \bibinfo
  {author} {\bibfnamefont {H.}~\bibnamefont {Pichler}}, \ and\ \bibinfo
  {author} {\bibfnamefont {M.~D.}\ \bibnamefont {Lukin}},\ }\href {\doibase
  10.1103/PhysRevLett.122.040603} {\bibfield  {journal} {\bibinfo  {journal}
  {Phys. Rev. Lett.}\ }\textbf {\bibinfo {volume} {122}},\ \bibinfo {pages}
  {040603} (\bibinfo {year} {2019})}\BibitemShut {NoStop}%
\bibitem [{\citenamefont {Prosen}\ and\ \citenamefont
  {{\v{Z}}unkovi{\v{c}}}(2013)}]{Bojan}%
  \BibitemOpen
  \bibfield  {author} {\bibinfo {author} {\bibfnamefont {T.}~\bibnamefont
  {Prosen}}\ and\ \bibinfo {author} {\bibfnamefont {B.}~\bibnamefont
  {{\v{Z}}unkovi{\v{c}}}},\ }\href {\doibase 10.1103/physrevlett.111.040602}
  {\bibfield  {journal} {\bibinfo  {journal} {Physical Review Letters}\
  }\textbf {\bibinfo {volume} {111}} (\bibinfo {year} {2013}),\
  10.1103/physrevlett.111.040602}\BibitemShut {NoStop}%
\bibitem [{\citenamefont {Das}\ \emph {et~al.}(2019)\citenamefont {Das},
  \citenamefont {Kulkarni}, \citenamefont {Spohn},\ and\ \citenamefont
  {Dhar}}]{PhysRevE.100.042116}%
  \BibitemOpen
  \bibfield  {author} {\bibinfo {author} {\bibfnamefont {A.}~\bibnamefont
  {Das}}, \bibinfo {author} {\bibfnamefont {M.}~\bibnamefont {Kulkarni}},
  \bibinfo {author} {\bibfnamefont {H.}~\bibnamefont {Spohn}}, \ and\ \bibinfo
  {author} {\bibfnamefont {A.}~\bibnamefont {Dhar}},\ }\href {\doibase
  10.1103/PhysRevE.100.042116} {\bibfield  {journal} {\bibinfo  {journal}
  {Phys. Rev. E}\ }\textbf {\bibinfo {volume} {100}},\ \bibinfo {pages}
  {042116} (\bibinfo {year} {2019})}\BibitemShut {NoStop}%
\bibitem [{ITe()}]{ITensor}%
  \BibitemOpen
  \href@noop {} {\bibinfo  {journal} {\mbox{ITensor Library} (version 2.0.11)
  http://itensor.org}\ }\BibitemShut {NoStop}%
\end{thebibliography}%

\end{document}